\documentstyle[nato,epsfig]{crckapb}

\begin{opening}
\title{SHOT NOISE OF MESOSCOPIC NS STRUCTURES :\protect\\
 THE ROLE OF ANDREEV REFLECTION}
\author{B. REULET$^{1,2}$}
\author{D.E. PROBER$^1$}
\institute{$^1$ Departments of Applied Physics and Physics, Yale University\\
New Haven CT 06520-8284, USA}
\institute{$^2$ Laboratoire de Physique des Solides, associ\'e au CNRS\\
b\^atiment 510, Universit\'e Paris-Sud\\
91405 ORSAY Cedex, France}
\author{W. BELZIG$^3$}
\institute{$^3$ Department of Physics and Astronomy, University of Basel\\
Klingelbergstr. 82, 4056 Basel, Switzerland}
\runningtitle{SHOT NOISE OF MESOSCOPIC NS STRUCTURES}
\end{opening}

\begin{document}

\section{Introduction}

When a mesoscopic wire made of normal metal N is in contact with a superconducting reservoir S, Andreev reflection (AR) occurs \cite{Andreev}. This affects the electronic properties of the wire \cite{NSreview}.
In this article we address both experimentally and theoretically the following question: how does Andreev reflection  manifest itself in shot noise measurements, and what physics can we deduce from such measurements ? Our discussion will rely on high frequency measurements performed on various NS structures, that can be found in refs. \cite{Schoelkopf97,Kozhevnikov00,Reulet02}.

The shot noise is a direct consequence of the granularity of the  electric charge. Even though an electric current exists only if there are free charged carriers, the value of this charge does not directly affect how much current flows through a sample when biased at a finite voltage. One has to investigate the fluctuations of the current to determine the charge of the carriers. As a consequence the measurement of the shot noise offers direct access to the elementary excitations of any system, through the determination of their effective charge. An effective charge different from $1e$ directly reflects how, due to their interactions, the electrons are correlated, as in a superconductor, a 2D electron gas in the fractionnal quantum Hall regime or a 1D Luttinger liquid. 

In an NS system, electrons can enter the superconductor only in pairs. Thus, the elementary charge participating in the electric current is no longer $e$, but $2e$. Thus, a naive expectation is that the shot noise should be doubled in the presence of an NS interface, as compared to the case of a normal metal. In the following we explore this expectation, to see how Andreev reflection affects current noise. In particular we show how a deviation of the effective charge from $2e$ reflects the existence of correlations among \emph{pairs} of electrons. 

In equilibrium the current noise power $S_I$ is given by the Johnson-Nyquist formula, which relates the current fluctuations to the conductance of the sample $G$: $S_I(V=0,T)=4k_BTG$, where $T$ is the electron temperature. Whether the fluctuating current is made of single electrons or pairs affects equilibrium noise only through the conductance, which may depend on the state (N or S) of the metal. 
This picture is valid at low frequency ($\hbar\omega\ll k_BT$). It is the 'classical' regime. At finite frequency $\omega$ a quantum mechanical treatment of noise is necessary. The general expression for the equilibrium current noise is given in terms of the reduced frequency $w=\hbar\omega/(2k_BT)$ by \cite{Kogan}:
\begin{equation}
S_I(V=0,T,\omega)=4k_BTG(T,\omega)g(w)
\label{SIomegaV0}
\end{equation}
where the function $g(x)$ is defined as: $g(x)=x\;\textrm{coth}\;x$.
As in the classical regime, all the physics is contained in the conductance $G$, but here $G$ is the real part of the complex and frequency dependent admittance of the sample. The $g(w)$ term accounts for statistical distribution of an excitation of energy $\hbar\omega$ at thermal equilibrium. Through $g(w)$ the finite frequency adds an energy scale at which the classical-to-quantum noise crossover takes place: $\hbar\omega=2k_BT$, or $w=1$.

Since we are interested here in shot noise, let us discuss first the case of the normal tunnel junction. The shot noise of the tunnel junction at low frequency is given by \cite{tunnel}:
\begin{equation}
S_I^{tunnel}(V,\omega=0,T)=4k_BTGg(v)
\end{equation}
with the reduced voltage $v=qV/(2k_BT)$. Here $g(v)$ interpolates between Johnson noise ($g(0)=1$) and shot noise ($g(v\gg1)=v$). Hence the charge $q$ appears at two levels: it can be measured through the equilibrium-to-shot noise crossover occuring at $qV=2k_BT$, or through the magnitude of the noise at high voltage, such that $S_I^{tunnel}=2qI$.
At finite frequency, the noise emitted by a tunnel junction is:
\begin{equation}
S_I^{tunnel}(V,\omega,T)=2k_BTG(g(v+w)+g(v-w))
\end{equation}
Thus, a finite frequency investigation offers another way to measure $q$, through the classical-to-quantum noise crossover occuring at $qV=\hbar\omega$.

In the case of a diffusive mesoscopic wire, the shot noise is reduced by the disorder, but the principles above are still valid. Thus, AR should show up in the \emph{shape} of the noise spectrum as a doubling of the classical-to-quantum noise crossover frequency, occuring at $\hbar\omega=2eV$. To perform such a measurement as a function of frequency, one needs to have a precise knowledge of $G(\omega)$ and of the frequency response of the experimental setup at high frequencies. (For  example, $T=100$mK corresponds to $V=8.6\mu$V and $\omega/2\pi=2.1$GHz.) As a consequence, the direct measurement of the frequency dependence of the noise spectrum at fixed voltage has never been accomplished. The noise measured in a narrow frequency band is expected to show the same crossover, but as a function of the applied voltage. This is a much easier (but still difficult) experiment, which we shall report in section \ref{section_freqdep}. The measurement has been performed so far on a normal metal wire, but could also be carried out on an NS sample. 
In Section \ref{section_PAN} we report measurements of photon assisted noise in an NS wire, which provide an alternative to the measurement of the crossover frequency. In this experiment a high frequency excitation is applied to the sample. The shot noise develops features as a function of voltage $V$ each time $qV$ is a multiple of the energy of the incident photons $\hbar\omega$. For the NS wire, $q=2e$.

The discussion above treats the consequences of the AR due to the energy $qV$, as compared to $k_BT$ or $\hbar\omega$. It investigates the effect of AR on the \emph{distribution statistics} of the electronic excitations (which involve pairs of electrons) rather than the \emph{effective charge} they carry. Specifically , the phenomena that have been measured and discussed above are related to steps in the distribution function, as will be discussed in section \ref{section_theory}. A better measurement of
 the effective charge is in the fully developed shot noise regime ($eV\gg k_BT$). Here the noise is determined by the fact that electrons are paired and also by the interferences  and the correlations that can exist among pairs. In section \ref{section_interf} we report measurements of phase dependent shot noise in an Andreev interferometer, which point out such a sensitivity of the effective charge to pair correlations. Section \ref{section_theory} is devoted to the theoretical investigation of the effective charge deduced from the shot noise.
Section \ref{section_exp} contains information about sample preparation and experimental setup, common to the experiments reported in subsequent sections.

\section{Experimental considerations}
\label{section_exp}

The experiments we report in the next sections use samples prepared with similar methods, and measured with similar detection schemes. Each sample consists of a metallic wire or loop between two metallic reservoirs, either normal or superconducting. The measurements are performed through contacts to the two reservoirs. This allows dc characterization of the sample (two-contact differential resistance $R_{diff}=dV/dI$, measured at $\sim200$Hz) as well as high frequency measurements. Even for the measurement of the effective charge of the Andreev interferometer, which does not intrinsically call for the use of rf techniques, high frequency measurements have been chosen for their extremely high sensitivity (the signal-to-noise ratio is proportionnal to the square root of the  bandwidth times the integration time of each measurement).

The measurements were performed in a dilution refrigerator at a mixing chamber 
temperature $T\sim~50$~mK. At low temperature the electron energy relaxation is dominated by electron-electron interactions~\cite{Altshuler:1982} and the associated inelastic length $L_{ee}$ is larger than $L$, so the transport in the device is elastic. We have not conducted weak localization measurements on these samples. From these, the phase coherence length $L_\varphi$ could have been extracted ($L_{ee}$ and $L_\varphi$ coincide if the dominant phase relaxation mechanism is electron-electron interaction, which is likely in our samples, otherwise $L_{ee}>L_\varphi$) \cite{Bergman}. Nevertheless, we observe significant harmonic content of the $R$ vs.~flux curve of the interferometer (data not shown). The n$^{th}$ harmonics decays as exp$(-nL/L^*(T))$, where $L$ is the distance between the two reservoirs. The empirical characteristic length $L^*$ includes phase beaking mechanisms ($L_\varphi$) as well as thermal averaging (usually described by the thermal length $L_T=(\hbar D/k_BT)^{1/2}$). We obtain $L^*(T=50$mK$)=800$nm of the order of $L_T$, which implies that $L_\varphi\gg L_T$, i.e., $L_\varphi$ is much larger than the sample size $L=540$nm. Other measurements on a longer interferometer ($L\sim1\mu$m) also gave evidence that $L_{ee}>L$ in that device. The wires described in sections \ref{section_freqdep} and \ref{section_PAN} are shorter and thus are also likely in the regime $L<L_\varphi$.

\subsection{Sample preparation}

The samples studied have been patterned by e-beam lithography. All are made of thin ($10$nm) evaporated gold wires between thick metallic reservoirs. The N wire and N reservoirs are deposited using a double angle evaporation technique \cite{FultonDolan} in a single vacuum pump down. Sputtered Nb is used for the thick ($80$ nm) S reservoirs. The transparency of the NS interface has been achieved by ion beam cleaning before Nb deposition. We estimate that the interface resistance is less than $1/10$ of the wire resistance. Theoretical calculations which consider this extra resistance show that its effect is negligible.
The Au wires have a temperature independent sheet resistance of the order of $\sim10~\Omega$ per square. The Au reservoirs are $70$~nm thick and have a sheet resistance of less than $\sim0.5~\Omega$ per square.

\begin{figure}[t]
\centerline{\psfig{height=8cm,figure=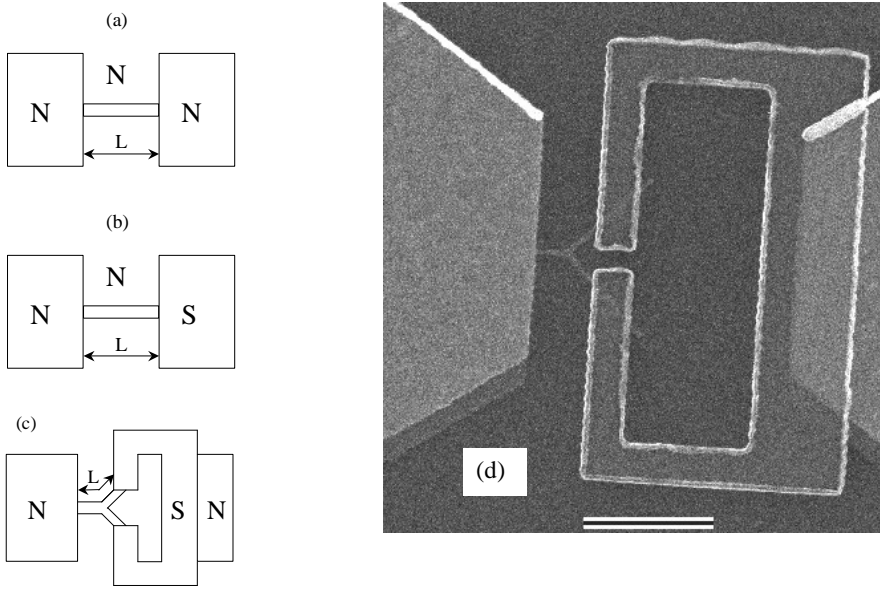}}
\caption{Schematics of the different samples that have been measured. (a) N wire between N reservoirs ($L=200$nm, $D=40$ cm$^2/$s). (b) N wire between N and S reservoirs ($L=280$nm, $D=30$cm$^2/$s). (c) Andreev interferometer ($L=540$nm, $D=33$cm$^2/$s). (d) SEM picture of the Andreev interferometer; N contact on right not shown. The scale bar corresponds to $1\mu$m.}
\label{fig_samples}
\end{figure}

\subsection{Experimental setup for high frequency noise measurements}

\begin{figure}[t]
\centerline{\psfig{height=8cm,figure=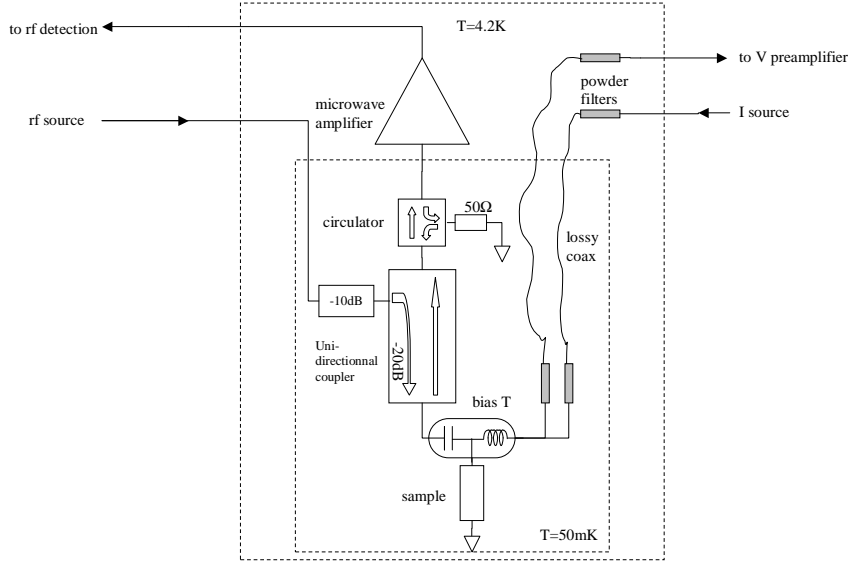}}
\caption{ Experimental setup used for high frequency measurements. The inner dotted line correspond to the mixing chamber at $T=50$mK, the outer one to the He bath or the 4K stage of the dilution refrigerator. The position of the coupler and the circulator are those used for the measurement performed on the Andreev interferometer only.
\label{fig_exp}}
\end{figure}

The experimental setup we used to perform the noise measurements on the Andreev interferometer is depicted in fig. \ref{fig_exp}. The current fluctuations $S_I$ in the sample are measured in a frequency band $\Delta f$ from $1.25$ to $1.75$GHz using an impedance matched cryogenic HEMT amplifier. The noise emitted by the sample passes through a cold circulator, employed to isolate the sample from amplifier emissions. It is then amplified by the cryogenic amplifier and rectified at room temperature after further amplification. The detected power is thus given by $P_{det}=G\Delta f(k_BT_{out}+k_BT_A)$ where $G$ is the gain of the amplification chain, $T_A\sim6.5$K is the noise temperature of the amplifier, and $T_{out}$ is the effective temperature corresponding to the noise power coming from the sample ($T_{out}=0.04-0.6$K for $V=0-150\;\mu$V for the case of an NS interface). We determine $G\Delta f$ and $T_A$ by measuring the sample's Johnson noise vs.~temperature at $V=0$ and its shot noise at $eV\gg(k_BT,E_C)$. ($E_c=\hbar D/L^2$ is the Thouless energy of a diffusive wire of length $L$; it is the energy corresponding to the inverse of the diffusion time along the wire).
We modulate the current through the sample to suppress the contribution of $T_A$. We measure $dP_{det}/dI$. This gives $dT_{out}/dI$. $T_{out}$ is related to the noise emitted by the sample through:
\begin{equation}
\label{Tout}
T_{out}=(1-|\Gamma|^2)T_N+|\Gamma|^2T_{in}
\end{equation}
and $S_I$ is given by $S_I=4k_BT_N\textrm{Re}Z_{diff}^{-1}$. Here $T_N$ is the sample's noise temperature, $Z_{diff}$ is the complex differential impedance of the sample at the measurement frequency, $\Gamma$ is the amplitude reflection coefficient of the sample and $T_{in}$ the external noise incoming to the sample. In eq. (\ref{Tout}), the first term on the right represents the noise emitted by the sample which is coupled to the amplifier. The second term represents the external noise the sample reflects. 

In order to determine $S_I$ at finite frequency, it is necessary to know both $Z_{diff}$ and $\Gamma$ at the measurement frequency. $Z_{diff}$ is deduced from the measurement of $\Gamma$ through:
\begin{equation}
\Gamma(\omega)=\frac{Z_{diff}(\omega)-Z_0(\omega)}{Z_{diff}(\omega)+Z_0(\omega)}
\end{equation}
where $Z_0$ the impedance of the measurement apparatus.
$Z_0$ is ideally real and equal to $50\Omega$. In practice it has an imaginary part and is frequency dependent (due to finite return loss of the amplifier or isolator, parasitic capacitance in parallel with the sample, inductance of the wire bond, etc.). Thus a careful calibration is necessary to have a reliable measurement of $Z_{diff}$\cite{Pieper}.
However, since our samples have a resistance close to $50\Omega$, the amplitude of $|\Gamma|^2$ is of the order of a few percent, and can be neglected in the first term of eq. (\ref{Tout}). This is not always the case for the second term. In the measurement performed on the normal wire (section \ref{section_freqdep}), no circulator was used.  $T_{in}\sim30K\gg T_N$ for this broadband (20 GHz) amplifier. The impedance of the sample (a very short gold wire) is voltage and frequency independent, so that the noise reflected by the sample adds up to the total as a voltage independent (but frequency dependent) constant. ($T_{in}$ depends on frequency because the amplifier emission does). For the NS wire (section \ref{section_PAN}), a circulator placed in liquid helium has been used. The circulator attenuates by 20dB the noise emitted by the narrowband amplifier towards the sample ($T_{emit}\sim2$K). In that case $T_{in}$ is equal to the temperature $T=4$K of the $50\Omega$ termination of the circulator. $T_{in}=(T_{emit}/100)+4K>T_{emit}$, but $T_{emit}$ drifts, which adds drift to the measurement. For this experiment one was interested only in the voltage dependence of the features of $S_I$, the corrections due to $T_{in}$ were not significant.

\begin{figure}[t]
\centerline{\psfig{width=0.9\textwidth,figure=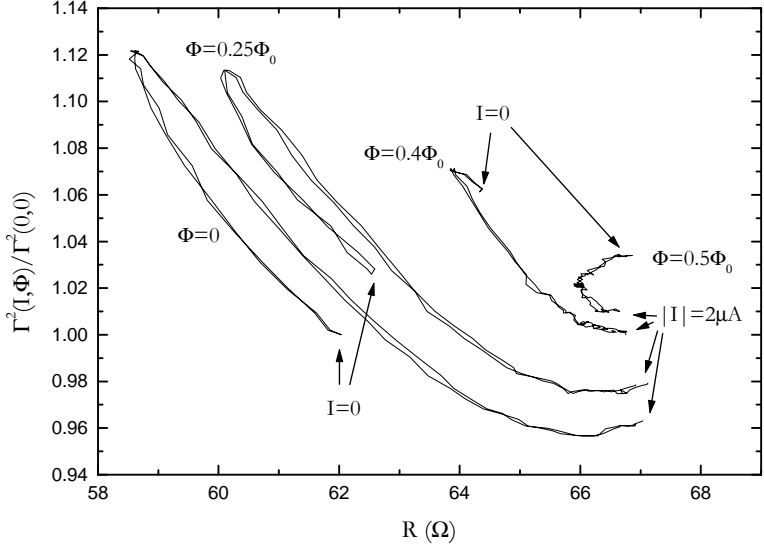}}
\caption{ Measurement of the variations of the power reflection coefficient $|\Gamma|^2$ vs. dc resistance of the Andreev interferometer. Each curve corresponds to a fixed magnetic flux and a varying current, between $-2\mu$A and $2\mu$A. The reflection coefficient has been arbitrarily rescaled to its value at $V=0$,$\Phi=0$. Conclusions are given in the text.
\label{fig_Gamma}}
\end{figure}

For the precise measurement of the effective charge in the Andreev interferometer, the magnitude of $S_I$ is of interest; it is thus crucial for $T_{in}$ to be minimized and stable. For this experiment we therefore placed the circulator at $T=50$mK. We also measured the variations of $|\Gamma|^2$ by sending white noise to the sample through the unidirectionnal coupler (see fig. \ref{fig_exp}) and detecting the change in the noise power. This determines the relative variation  of the amplitude of the reflection coefficient (as a function of voltage or magnetic flux) over the bandwidth used for the noise measurement. We draw the following conclusions: i) the variations of $|\Gamma|^2$ are small enough to be neglected, allowing us to take $\Gamma=0$ in the data analysis. This is confirmed by the fact that at $V=0$, where $T_N=T$, $R_{diff}$ and $\Gamma^2$ are flux dependent; yet $T_{out}$ at $V=0$,  does not depend on the flux (see eq.(\ref{Tout})); ii) the impedance of the sample at the measurement frequency is different from its dc value. This conclusion is seen from the following : if $Z_{diff}(\omega)$ were equal to its dc value $R_{diff}=dV/dI$, then whatever $Z_0(\omega)$ is (i.e. whatever the imperfections of the experiment are), $|\Gamma|^2$ plotted as a function of $R_{diff}$ should collapse into a single curve for all the values of flux and voltage. As shown on fig. \ref{fig_Gamma}, this doesn't occur. This means that the impedance of the sample is not equal to $R_{diff}=dV/dI$ measured at low frequency. It has an flux- or voltage-dependent imaginary part, or its real part is not simply proportionnal to $R_{diff}$. This observation deserves more study, through measurements of the amplitude and phase of the reflection coefficient as a function of flux, voltage and frequency.
However, this effect is small, and for our present study of noise, we simply use $R_{diff}$ to determine $S_I$ from $T_N$.
This is also justified by the fact that in our short phase-coherent samples transport is elastic. Thus, the ac conductance is given by the dc $I(V)$ characteristics shifted by $\pm\hbar\omega/e\approx\pm6\mu$V~\cite{tunnel}. Since the characteristic scale for changes of $dV/dI$ is $\sim30\mu$V, finite frequency corrections to  $R_{diff}$ should be  small.

In the experiment described in section \ref{section_freqdep}, the noise needs to be measured over a broad frequency range. Thus, a broadband ($1-20$ GHz) cryogenic amplifier has been used, even though it has a higher noise temperature ($T_A\sim100$K) than a narrow band amplifier. Also a circulator cannot be used. The noise at different frequencies is obtained by measuring the low frequency noise power after heterodyne mixing (at room temperature) the amplified signal from the sample against a variable frequency oscillator.

\section{Measurement of the frequency dependence of the shot noise in a diffusive N wire}
\label{section_freqdep}

\begin{figure}[t]
\centerline{\psfig{width=0.95\textwidth,figure=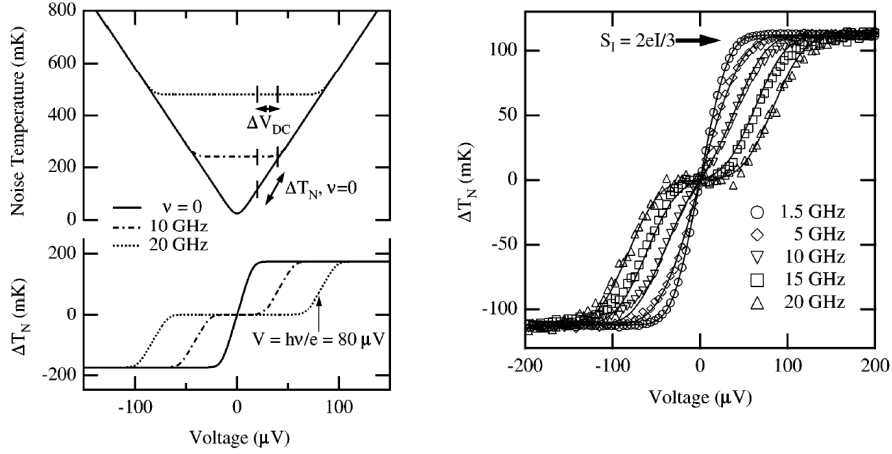}}
\caption{ Left: predicted dc bias voltage dependence of noise (top)
for three frequencies at a bath temperature of 25 mK. The
current spectral density predicted by eq. (\ref{eq_SIN}) has been converted to an equivalent noise temperature
$T_N$ through the relation $T_N=S_I/(4k_BG)$. Note that the noise is
independent of bias voltage for $e|V|<\hbar\omega$. The bias voltage
modulation technique we employed is shown schematically and the expected
differential noise $\Delta T_N$ for a $30\mu$V p.p. modulation is also
displayed (bottom).
Right: measured differential noise for frequencies of $\nu=\omega/2\pi=$1.5,
5, 10, 15, and 20 GHz, with mixing chamber temperature of
40 mK. Solid lines show the predictions of eq. (\ref{eq_SIN}) for an
electron temperature of 100 mK, and accounting for the voltage
modulation of 30 $\mu$V p.p.}
\label{figs97}
\end{figure}

In this section we report measurements of the frequency dependence of the out-of-equilibrium ($V\neq0$) noise in a normal metal wire between two N reservoirs (see fig. \ref{fig_samples}(a)) \cite{Schoelkopf97}.
In such a system the current noise at finite frequency is given by \cite{BuBlan}:

\begin{equation}
\label{eq_SIN}
S_I(\omega,T,V)=4k_BTG\left( \frac\eta2\left(g(v+w)+g(v-w)\right)+(1-\eta)g(w) \right)
\end{equation}
The $\eta=1/3$ factor corresponds to the shot noise reduction due to disorder (Fano factor)\cite{BuBlan}.
The shot noise corresponds to the $v$ terms whereas equilibrium noise is given by $v=0$. As shown by eq. (\ref{eq_SIN}), the total noise is \emph{not} given by the sum of equilibrium noise (classical Johnson or quantum) and shot noise, even at zero frequency ($w=0$).
The unusual nature of this superposition can be emphasized
by examining the fluctuations predicted by eq. (\ref{eq_SIN})
as a function of voltage for different frequencies (see
fig. \ref{figs97} left). At zero frequency (full line), there is a transition
from Johnson noise to the linearly rising shot noise
at $eV\sim k_BT\sim2\;\mu$eV. At $20$ GHz (dotted line), the
fluctuations are dominated by quantum noise and do not
increase from their value at equilibrium, until the voltage
$V_c=\hbar\omega/e\sim80\mu$V is exceeded, even though the condition
$eV>k_BT$ is fulfilled and the low-frequency fluctuations
(solid line) are increasing rapidly. Only above $V_c$ does the dc voltage provide enough energy to increase the emitted photon noise.

The differential noise theoretically expected for
the diffusive conductor, under the example conditions of
$T=25$ mK, and a small ($\Delta V=30 \mu$V p.p.) square-wave voltage
modulation, are shown in the bottom left of fig. \ref{figs97}. The square-wave modulation contributes significantly to the width of the rises. 
The noise measurements are reported as a variation, $\Delta T_N$, of the sample's noise temperature due to the voltage modulation $\Delta V$. The
measured values of $\Delta T_N$ for frequencies of 1.5, 5, 10,
15, and 20 GHz, taken at a mixing chamber temperature
of $T=40$ mK, are shown in fig. \ref{figs97} right. While we see that $\Delta T_N$
for the low-frequency noise (circles) changes rapidly with
voltage, approaching its linear asymptote at voltages
only a few times $k_BT/e$, the curves become
successively broader for increasing frequency. The noise
for the highest frequencies has a clearly different shape,
displaying the expected plateau around $V=0$. Also shown in fig. \ref{figs97} right (full
lines) are theoretical curves based on eq. (\ref{eq_SIN}), accounting
for the finite voltage difference used, and for an electron
temperature of 100 mK. The asymptotic value of
$\Delta T_N$ has been arbitrarily scaled (since the frequency dependent
system gain is not known to better than about
$30\%$) to be 112 mK for each frequency, corresponding to
the expected reduction factor of $\eta=1/3$ (see eq.(\ref{eq_SIN})). Note that such a reduction of the shot noise could also be attributed to the heating of the electrons (hot electron regime)\cite{Steinbach}. See ref.\cite{Schoelkopf97} for a detailed discussion of this possibility.

Though it has not been measured, in an NS geometry the same qualitative behaviour is expected, with the scale $eV$ replaced by $2eV$. For the NS system  there might be corrections at a frequency of the order of the Thouless energy, in the same way there is a signature at $\sim4E_c$ in the voltage dependence of the noise measured at low frequency (see section \ref{section_interf}).

\section{Observation of photon assited noise in an NS wire}
\label{section_PAN}

In this section we report measurements of low frequency noise (measured in the bandwidth $1.25-1.75$GHz) emitted by an NS sample which experiences both a dc and an ac bias \cite{Kozhevnikov00}. The sample is a thin Au wire between N and S reservoirs, as depicted in fig. \ref{fig_samples}(b). In the presence of an ac excitation at frequency $\omega$, the noise emitted by a diffusive wire is:
\begin{equation}
S_I(V,T)=4k_BTG\left((1-\eta) + 2\eta \sum_{n=-\infty}^{+\infty}J_n^2(\alpha)g(v+nw)\right)
\end{equation}
where $\alpha=2eV_{ac}/(\hbar\omega)$ is a dimensionless parameter measuring the amplitude of the ac excitation voltage (note that $\omega$ denotes here the frequency of the ac bias; the the frequency at which the noise is measured is considered to be dc). This formula is valid at low ($E\ll E_c$) and high ($E\gg E_c$) energy, where $G$ is voltage- and frequency-independent. In between, the energy dependence of the Andreev process may give rise to corrections, as is the case for the conductance and for the effective charge (see section \ref{section_theory}).

\begin{figure}
\centerline{\psfig{width=0.9\textwidth,figure=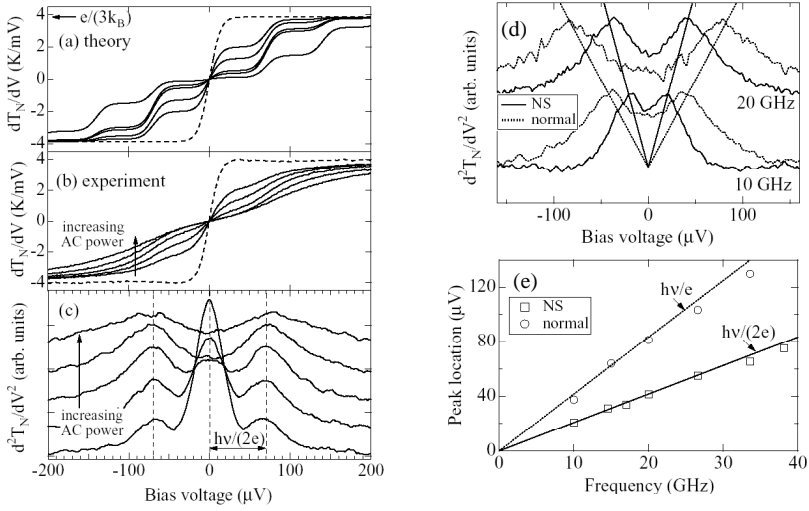}}
\caption{ Predicted and observed shot noise of an N-S device
vs. bias voltage without ac bias and at different powers
of ac excitation at 34 GHz: (a) theory for $dT_N/dV$ at
$T=100$ mK with no ac (dashed line) and with ac excitation
at $\alpha= 1.1, 1.4, 1.7, 2.2, 2.8$ (solid lines); (b) experimentally
measured $dT_N/dV$ with no ac bias (dashed line) and with ac
excitation powers differing by 2 dB and corresponding to the
above values of $\alpha$ (solid lines); (c) $d^2T_N/dV^2$ obtained by
numerical differentiation of data in (b).
(d) $d^2T_N/dV^2$ vs. bias voltage at $B=0$ (solid lines)
and at $B=5$ T (dotted lines) with ac excitation at $\hbar\omega/2\pi=10$ and $20$ GHz. The curves are offset vertically by an amount proportional to frequency. The solid straight lines mark the expected
peak locations for the N-S case (at $B=0$): $V_{peak}=\hbar\omega/(2e)$;
the dotted straight lines mark the expected peak locations for
a normal device (at $B=5$ T): $V_{peak}=\hbar\omega/e$; (e) peak location
vs. frequency for $B=0$ and $B=5$ T; the solid and the
dotted straight lines are $V_{peak}=\hbar\omega/(2e)$ and $V_{peak}=\hbar\omega/e$, respectively.}
\label{figs00}
\end{figure}

In the absence of ac excitation, the measured differential
noise $dT_N/dV$ vs. bias voltage for the N-S device is, within $5\%$, twice as big as that measured
when the device is driven normal by a magnetic field of 5 T \cite{KozJLTP}. This is a measure of the doubling of the effective charge, which will be discussed more in section \ref{section_interf}.
We now turn to the noise measured in the presence of an ac excitation.
If the transport is still elastic in the presence of ac excitation, the shot noise is expected to develop features at bias
voltages such that $qV=\pm n\hbar\omega$. The location of these features
should be independent of ac power. In contrast, if the transport
is inelastic, no photon-assisted features should occur.
The derivative of the noise vs. bias voltage was measured
with ac excitation at $\omega/2\pi=34$ GHz, at different levels of
ac power. Figures \ref{figs00}(a) and (b) show the predicted and
observed derivative of the noise temperature vs. dc bias
voltage for several levels of ac power. To see the features
more clearly, we plot in fig. \ref{figs00}(c) the second derivative
$d^2T_N/dV^2$ obtained by numerical differentiation of
the experimental data. With no ac excitation, $d^2T_N/dV^2$
has a peak at $V=0$. With ac excitation, the sidebands
of this peak are clearly evident at $V=\pm n\hbar\omega/(2e)$. The
sideband locations are power independent, which further
argues that the structure is due to a photon-assisted
process. The magnitude of $d^2T_N/dV^2$ at $V=0$ displays
oscillatory (roughly $\sim J^2(\alpha)$)  behavior vs. ac excitation
amplitude (not shown), which is another hallmark of a photon-assisted process.
We note that photon assisted processes are seen clearly in SIS tunnel junctions\cite{Tucker}. The features there are in the quasiparticle current, so the charge involved in that case is $1e$. They are centered at the gap voltage $V=2\Delta/e$, since at low temperature pair breaking must occur for a quasiparticle to tunnel.

The most convincing evidence of the photon-assisted nature
of the observed effects is the dependence of the voltage
location of the sideband peak on the frequency of the
ac excitation. Measurements of the shot
noise were made at several different frequencies of ac bias, both in
zero magnetic field and at $B=5$ T, for which the Nb reservoir is driven normal. Figure \ref{figs00}(d) shows
the second derivative of the shot noise power vs. bias voltage
for $\hbar\omega/2\pi=10$ and $20$ GHz at $B=0$ (solid lines) and for the
same device at $B=5$ T (dotted lines), where the sample is driven normal. The solid and dotted
straight lines are the expected peak positions for the
N-S and normal cases, respectively. The peak locations
clearly follow the theoretical predictions $V=\hbar\omega/q$
with $q=2e$ in the case of the N-S device and $q=e$ in the case
of the device driven normal. Figure \ref{figs00}(e) shows the peak
locations for a number of different ac excitation frequencies
at $B=0$ and $B=5$ T. The solid and dotted lines are
theoretical predictions with no adjustable parameters.

\section{Measurement of the phase dependent effective charge in an Andreev interferometer}
\label{section_interf}

In this section we show precise measurements of the effective charge $q_{eff}$, and how deviations from $q_{eff}=2e$ are phase sensitive. We relate these deviations to correlations of the charge transfer during Andreev reflection \cite{Reulet02}.
After an Andreev process, the reflected hole carries information about the phase of the superconducting order parameter of the S reservoir at the N-S interface. When two S reservoirs are connected to the same phase-coherent normal region, a phase gradient develops along the normal metal, resulting in phase-dependent properties. In an Andreev interferometer - a device containing a mesoscopic multiterminal normal region with a (macroscopic) superconducting loop, all the electronic properties are periodic with the magnetic flux $\Phi$ enclosed by the loop, with a period of the flux quantum, $\Phi_0 = h/(2e)$. The sample used for this experiment is depicted in fig. \ref{fig_samples}(c) and (d).

\begin{figure}
\centerline{\psfig{height=11cm,figure=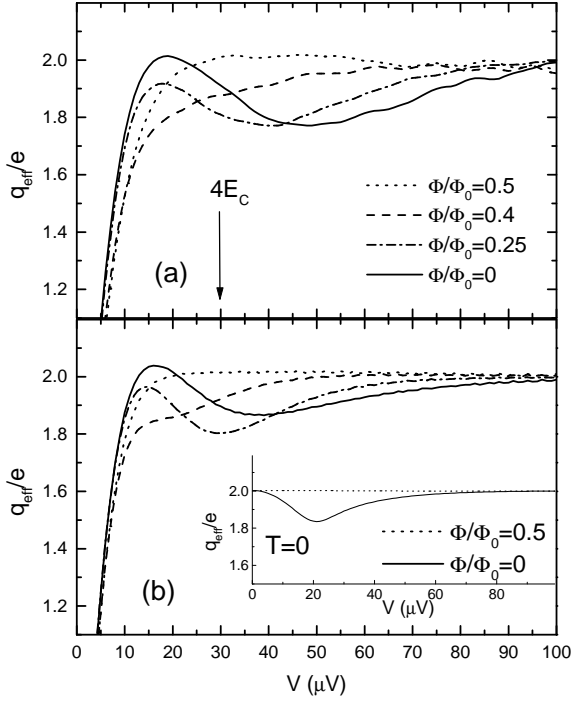}}
\caption{(a) Experimentally measured effective charge $q_{eff}$ for several values of magnetic flux. (b) Theoretical predictions for $E_C=7.5\mu$eV and $T=43$mK. The dip in $q_{eff}$ is predicted to occur at $\sim4E_c$. The inset shows the theory for $\Phi=0$ and $\Phi=\Phi_0/2$ at $T=0$. Note that our definition $E_c=\hbar D/L^2$ uses $L$ for the full length of the normal region, and thus differs from the definition in Ref. \protect\cite{Reulet02}.}
\label{PRL02_fig3}
\end{figure}
 
From the noise measurements we deduce the effective charge, $q_{eff}=(3/2)(dS_I/dI)$; see fig. \ref{PRL02_fig3}(a). 
By considering $dS_I/dI$ rather than $dS_I/dV$ we eliminate the trivial effect of a non-linear $I(V)$ characteristic. At finite energy ($E>k_BT$) the effective charge reflects the charge transferred but also includes the effects of correlations in the transfer process.  The voltage dependence of $dS_I/dI$ yields information on energy-dependent correlations between charge transfers.
Figure \ref{PRL02_fig3}(b) gives the theory results based on full counting statistics. The inset shows the theory for $\Phi=0$ and $\Phi=\Phi_0/2$ at $T=0$.
The effective charge is seen in the theory to be independent of the phase difference
at bias voltages larger than $\sim100~\mu$V, with significant phase modulation of 
$q_{eff}$  in the bias voltage range $10-80~\mu$V. The maximum magnitude
of the observed dip of $q_{eff}$ vs. voltage is $\sim10$\%, and occurs for $\Phi\sim\Phi_0/4$. There is no dip for $\Phi=\Phi_0/2$. For $T=0$, $q_{eff}$ returns to $2e$ as $V\rightarrow0$ (see inset of fig.\ref{PRL02_fig3}). At finite temperature, $q_{eff}$ goes to zero for $eV\ll k_BT$. This is because $S_I$ reduces to Johnson noise at $V=0$. Thus, the decrease of $q_{eff}$ at finite temperature and at very low voltages is not related to Andreev physics. In contrast, the dip near $4E_c\sim30\;\mu$eV (with $E_c=\hbar D/L^2$) is due to the energy dependence of the Andreev processes.

The experimental results are in fairly good agreement with the theoretical predictions. As expected, there is no phase modulation of $q_{eff}$ at large energies $eV\gg E_c$, and here $q_{eff}=2e$. At $E\sim4E_c$, the effective charge is smaller for integer flux than for half-integer
flux. The non-trivial energy- and flux dependence predicted (crossings of the different curves) is seen in the experiment, though the agreement is not perfect. The magnitude of the dip of $q_{eff}$ in the data is also close to the theoretical prediction.

To understand the origin of the dip of $S_I$ seen for $\Phi=0$, we have also solved a generalized Boltzmann-Langevin (BL) equation. In such an approach correlations due to the superconductor enter only through the energy- and space-dependent conductivity, which gives $I(V,\Phi)$. Thus, the BL result is not complete, and we will compare its predictions to that of the full-counting-statistics theory, to help understand those predictions. At $T=0$, the BL result for all flux values is simply $S_I^{BL}=(2/3)2eI(V,\Phi)$, i.e., $q_{eff}=2e$ at all energies. This implies that the deviation of the effective charge from $2e$, measured and predicted by the full theory, must be due to fluctuation processes which are not related to single-particle scattering, on which the BL approach is based.  
We believe that the higher-order process which is responsible for the dip of $S_I$ is a two-pair correlation process.  At high energies ($E\gg E_c$) the electron-hole pair states have a length $\sim(\hbar D/E)^{1/2}$, shorter than $L$. This results in uncorrelated entry of pairs into the normal region.  For $E\sim E_C$ the pair size is larger, and the spatial overlap prevents fully random entry, suppressing $S_I$. Suppressed shot noise is a signature of anti-correlated charge entry~\cite{BuBlan}. At yet lower energies (at $T=0$) the effective charge is predicted to return to 2e; we do not yet have a physical interpretation of this. In any case, for the case of $\Phi=\Phi_0/2$, the dip of $q_{eff}$ is fully suppressed, according to the theory. This means that the phase gradients destroy the pair correlation effect responsible for the dip.  

\section{Theoretical approach to the effective charge}
\label{section_theory}

We now turn to our theoretical approach to current noise in mesoscopic
proximity effect structures. Our goal in this section is to explain the predictions and the meaning of the effective charge. We shall see that the 'dip' in the effective charge seen for the interferometer is also seen in wires, and arises from similar pair correlation effects. A characteristic feature of the NS
structures is that the phase coherent propagation of Andreev pairs in
the normal metal is influenced by the proximity effect. One consequence
is the so-called reentrant behaviour of the conductance of a normal
diffusive wire in good contact to a superconducting terminal \cite{NSreview}. The
conductance is enhanced at energies of the order of $\sim5E_c$.
At higher and lower energies the conductance approaches its
normal state value \cite{yuli:96}. In a fork geometry, as discussed in
the previous section, the proximity effect can be tuned by a phase
difference between the two superconducting terminals.

\begin{figure}[t]
    \centerline{\psfig{width=0.6\textwidth,figure=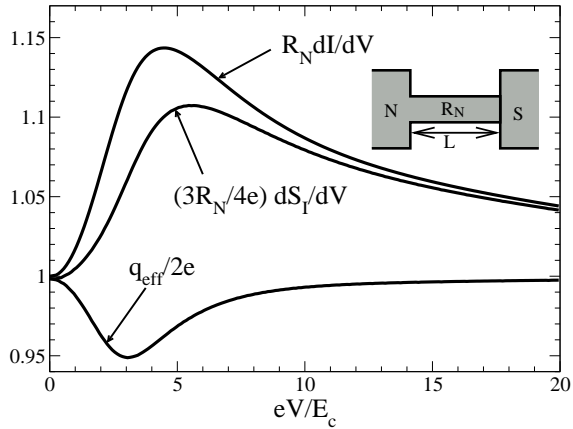}}
    \caption{ Transport characteristics of a proximity wire. The inset
      shows the layout. The relevant energy scale is the Thouless energy
      $E_c=\hbar D/L^2$. In the main plot the differential
      conductance, the differential noise $dS_I(V)/dV$, and the
      effective charge as defined in (\ref{eq:qeff-def}) at zero
      temperature are shown.}
    \label{fig:snwire}
\end{figure}

We are interested in the bias-voltage dependence of the current noise in
a diffusive NS structure.  There is, however, a simple energy dependence
resulting from the proximity-induced energy-dependent conductivity. In contrast, the
correlations of interest are not due to this energy-dependence of the conductivity. They can be
distinguished in the following way.  We note that for the average
transport properties, i.e. the differential conductance, the kinetic equation takes the very simple form
\cite{yuli:96,vzk,suplatrev,yuli:99-1}
\begin{equation}
  \label{eq:kinetic}
  \nabla \sigma({\mathbf x},E)\nabla f_T({\mathbf x},E)=0\,. 
\end{equation}
The energy- and space-dependent conductivity $\sigma({\mathbf x},E)$
includes the proximity effect, and
$f_T({\mathbf x},E)=1-f({\mathbf x},E)-f({\mathbf x},-E)$ is the symmetrized distribution function.
Due to the induced superconducting correlations, $\sigma$ is enhanced above its
normal state value $\sigma_N$ and its energy- and space dependence is
obtained from the spectral part of the Usadel equation \cite{yuli:96,usadel}.  For the geometry in fig.~\ref{fig:snwire}
the spectral conductance is
$G^{-1}(E)=\int dx/\sigma(x,E)$. Note that $G(E)\neq 0$ even for $E\ll \Delta$ as a
consequence of the proximity effect. The current for a
given bias voltage $V$ and temperature $T$ is then given by
\begin{equation}
  \label{eq:fork-current}
  I(V,T)=\frac{1}{2e}\int_{-\infty}^{+\infty} G(E) f_T^N(E,V,T) dE
\end{equation}
where $f_T^N(E,V,T)$ is the symmetrized distribution function in the normal
metal reservoir and we have accounted for the boundary condition
$f_T^S(E\ll\Delta,V,T)=0$ at the superconducting terminal.

The form of the kinetic equation (\ref{eq:kinetic}) suggests that
electrons and holes (i.e.  positive and negative energy quasiparticles)
obey \emph{independent} diffusion equations, which are only coupled
through the boundary condition $f_T^S=0$ at the superconducting
terminal. Thus, we may try to apply the semiclassical Boltzmann-Langevin
(BL) approach \cite{nagaev,nagbuett}. The only modification is that we
have to account for an energy- and space-dependent conductivity.  For
that purpose it is convenient to introduce the characteristic potential,
defined as the solution of the equation :
\begin{equation}
  \label{eq:charpot}
  \nabla \sigma(x,E)\nabla \nu(x,E)=0
\end{equation}
with the boundary condition that $\nu=1$ at the normal terminal and
$\nu=0$ at the superconducting terminal. The solution for our
quasi-one dimensional geometry is :
\begin{equation}
  \label{eq:charpotsol}
  \nu(x,E)=G(E)\int_x^L \frac{dx}{\sigma(E,x)}
\end{equation}
The current noise can then be expressed in the familiar form :
\begin{equation}
  \label{eq:bl-noise}
  S_I^{BL}(V)=4 \int dE \int dx \sigma(x,E) \left(\nabla \nu(x,E)\right)^2
  f(x,E)(1-f(x,E))
\end{equation}  
where the distribution function is given by $f(x,E)=\nu(x,E)f^N(E,V,T)$.
As a result we find for the noise at zero temperature (i.e.,
$f_T^N(E,V,T=0)= $ -sign$(eV)$ for $|E|<|eV|$ and zero otherwise) :
\begin{equation}
  \label{eq:bl-noiseresult}
  S_I^{BL}(V)=\frac{4e}{3}I(V)
\end{equation}
where the current is given by $I(V)=(1/2e)\int_{-eV}^{eV} dE G(E)$.
Thus, we find that the current noise depends in a non linear fashion on the voltage. The nonlinearity is given by the $I(V)$ characteristic.
However, this
dependence is in some sense trivial, since the only way the
electron-hole coherence enters is through the energy dependent
conductivity. 

We note that the doubling of the shot noise in comparison to the normal
case results from the \emph{energy integration} from $-eV$ to $+eV$,
instead of the interval between $0$ (i.e., the Fermi energy) and $eV$, as it would
be in the normal case. On the other hand, the average 'noisiness' (coming
from the spatial integral in eq.~(\ref{eq:bl-noise}) alone) is the
\emph{same} in the normal and the superconducting cases \cite{nagbuett}. Thus, the
doubling of the noise needs not be interpreted as a direct consequence of the doubled charge
transfer involved in an Andreev reflection process. It reflects
the particle-hole symmetry in the superconducting terminal. Nevertheless
we adopt below the notion of an effective charge, since it
is a convenient measure of the deviation from the independent electron
fluctuations.

These arguments can also be used to explain the experiments on
photon-assisted noise described previously. In the presence of an
ac voltage of frequency $\omega$ the electron distribution in the normal
terminal acquires side-bands, i.e., additional steps at
energies $\pm n\hbar\omega$. The noise of the diffusive wire depends
essentially on a superposition of left and right distribution functions
differing by the voltage $eV$ in the normal state and by $2eV$ in the
NS case. It is clear that the noise properties change as a function of
voltage, when sideband features match the other steps in the distribution
function. As a consequence, photon-assisted steps occur when the voltage
matches $n\hbar\omega/e$ in the normal case and $n\hbar\omega/2e$ in the
superconducting case. This is what is observed in the
experiments (see section \ref{section_PAN}).

A correct calculation of the noise requires that we go beyond the independent electron- and
hole\--fluc\-tuations in the Boltzmann-Langevin approach. This can be accessed by
the extended Green's function approach \cite{belzig:01-diff}. To
describe the fluctuations, we utilize the results of the full counting statistics and define an effective charge
\begin{equation}
  \label{eq:qeff-def}
  q_{eff}(V,T)=\frac{3}{2}\frac{\partial S_I(V,T)}{\partial I(V,T)}\,,
\end{equation}
which takes the value $2e$ for the Boltzmann-Langevin result (see eq.
(\ref{eq:bl-noiseresult})). It follows that the energy dependence of the
effective charge gives information about the correlated electron-hole
fluctuation processes.

In order to illustrate these considerations, we show in fig.~\ref{fig:snwire}
results for the conductance $dI/dV$, the differential noise $dS_I(V)/dV$, and
the effective charge at zero temperature for a one-dimensional diffusive
wire between a normal and a superconducting reservoir. Note: $q_{eff}$ is proportionnal to $dS_I/dI=(dS_I/dV)(dI/dV)^{-1}$. These results
were obtained by a numerical solution of the quantum-kinetic equation.
The differential conductance shows the well-known reentrance (peak) behaviour
\cite{yuli:96}. At low and high energy the conductance approaches the
normal state value $G_N$. At intermediate energies of the order of the
Thouless energy the conductance is enhanced by $\sim 15\%$ above $G_N$. A similar result is found for the interferometer when $\Phi=0$ \cite{Reulet02}.
We observe that the differential noise $dS/dV$ has a roughly similar energy
dependence\cite{belzig:01-diff}, although it is quantitatively
different. The deviation of the voltage-dependent effective charge from
$2e$ demonstrates the energy dependence of the higher order
correlations, which are not contained in the independent electron-hole
picture, the BL approach. Around $E\sim4E_c$ the effective charge is suppressed below $2e$,
showing that the higher order correlations result in a reduced noise in
comparison to the BL case of uncorrelated electron and hole fluctuations, for which $q_{eff}=2e$. As $V\rightarrow0$, $q_{eff}$ approaches $2e$ again.

The physics of this effective charge is clarified if we consider the full counting statistics
\cite{belzig:01-diff,belzig:01-super,levitov:96-coherent,yuli:99-annals},
instead of the current noise only. In full counting statistics, we
obtains the \emph{distribution of tranfered charges}, which clearly
contains direct information on both the statistics and the nature of the
charge carriers.  There it follows that \emph{all} charge transfers at
subgap-energies occur in units of $2e$ \cite{kmel}.  However, this does
not necessarily result in a doubled effective charge using our definition of $q_{eff}$. The effective charge also includes the effect of correlations between the different charge transfers. Our work in section \ref{section_interf} shows that these correlations are phase-dependent.Nevertheless, in the limit of \emph{uncorrelated} transfer of Andreev pairs, the
effective charge at $E\gg k_BT$ is simply $2e$.

\section{Conclusion}

In this paper we have discussed how Andreev Reflection affects shot noise of mesoscopic NS structures. High frequency measurements provide a very powerfull tool since they are very sensitive, as is essential to the measurement of the small voltage- and flux dependence of the effective charge, and these measurements access an energy domain in which interesting phenomena occur, when $\hbar\omega>eV,k_BT$. These techniques are very promising for investigation of even more subtle quantities like cross-correlations in the noise or higher moments of the current fluctuations \cite{ReuletS3}. The measurements have revealed the existence of correlations in pair charge transfers which are not accessible through conductance measurements. The full counting statistics method gives access to the full distribution of the charge transfers, and is essential for understanding the physics of such NS structures. This method sheds light on the correlations revealed by the shot noise, and allows as well the investigation of cross-correlations and higher moments.

\section*{Acknowledgements}

The authors thank R.J. Schoelkopf, A.A. Kozhevnikov, P.J. Burke and M.J. Rooks for collaboration on some of the experiments reported, and also M. Devoret and I. Siddiqi and Yu V. Nazarov for useful discussions. This work was supported by NSF DMR grant 0072022. The work of W.B. was supported by the Swiss NSF and the NCCR Nanoscience.

\end{document}